# Optical Wireless Hybrid Networks for 5G and Beyond Communications

Mostafa Zaman Chowdhury, Moh. Khalid Hasan, Md. Shahjalal, Md. Tanvir Hossan, and Yeong Min Jang
Dept. of Electronics Engineering, Kookmin University, Seoul, Korea
E-mail: mzaman@kookmin.ac.kr, khalidrahman45@ieee.org, mdshahjalal26@ieee.org, mthossan@ieee.org, yjang@kookmin.ac.kr

*Abstract*—The next 5$^{th}$ generation (5G) and above ultra-high speed, ultra-low latency, and extremely high reliable communication systems will consist of heterogeneous networks. These heterogeneous networks will consist not only radio frequency (RF) based systems but also optical wireless based systems. Hybrid architectures among different networks is an excellent approach for achieving the required level of service quality. In this paper, we provide the opportunities bring by hybrid systems considering RF as well as optical wireless based communication technologies. We also discuss about the key research direction of hybrid network systems.

## 1. Introduction

The communication era is shifting from 4$^{th}$ generation (4G) to 5$^{th}$ generation (5G). 5G and above communication systems will provide ultra-high level of service quality. Convergence of heterogeneous wireless networks is one of the most important issues for these communication systems. The heterogeneous networks consist of both radio frequency (RF) and optical wireless based networks.

Currently, RF based technologies are widely used for wireless connectivity. The use of this band is strictly regulated [1]. RF has limitations such as small spectrum band, regulation in the use of spectrum, sever interference between nearby RF access points (APs), etc. Optical wireless communication (OWC) networks [1]-[5], the complementary of RF based networks is considering as a promising solution. OWC refers to communication through propagation media of visible light (VL), infrared (IR), or ultraviolet (UV). The commonly used OWC technologies can be categorized into two groups (i) communication through visible light and (ii) communication through non-visible light. In 2$^{nd}$ group, illumination is not a concern at all. The most promising OWC technologies are visible light communication (VLC), light fidelity (LiFi), optical camera communication (OCC), and free space optical communication (FSOC) [1].

Each single network considering both RF and optical wireless based systems has its excellent features as well as limitations. The hybrid architecture consisting two or more networks overcomes the limitations of individual networks. This paper discusses the opportunities created by different hybrid communication systems. We also provide clear research direction for future hybrid network systems consisting of optical wireless networks.

The rest of the paper is organized as follows. Section 2 provides briefly overview of OWC and RF systems those are considering for hybrid network architectures. The hybrid RF/optical systems are discussed in Section 3. Section 4 deliberated solutions for optical/optical hybrid network systems. Section 5 gives research directions for such hybrid network systems. Finally, we conclude our paper in Section 6.

## 2. Networks to be Considered for Hybrid Systems

OWC technologies such as VLC, LiFi, OCC, and FSOC; and RF technologies such as WiFi, 4G/5G small-small cell [6], [7], macrocell, microwave-link are considered for hybrid systems in this paper. A brief description of these technologies are given in this section. Fig. 1 shows the basic connectivity of these technologies.

### 2.1 Optical wireless communication systems

Visible light communication [2], [8], [9]: It uses Light-emitting diode (LEDs) luminaires or laser diodes (LDs) as transmitters and photodetectors (PDs) as receivers. It uses only VL as the communication media. It can provide communication, illumination, and localization. A 100 Gbps data rate is already achieved using VLC [10].

Light fidelity [11], [12]: LiFi technology is similar of wireless fidelity (WiFi). It provides high-speed wireless connectivity along with illumination. It uses LEDs or LDs as transmitters and PDs as receivers, respectively. For the communication media, it uses VL in the forwarding path and VL/IR/UV as the returning path. It can provide communication, illumination, and localization.

Optical camera communication [1], [5]: OCC uses LEDs and a camera as the transmitter and receiver, respectively. It can use IR or VL as the communication media. This technology can provide better performance even in outdoor environments.

FSO communication (FSOC) [13]: FSOC uses LD and PD as transmitter and receiver, respectively. It can provide very long distance (more than 10,000 km) communication with very high data rate.

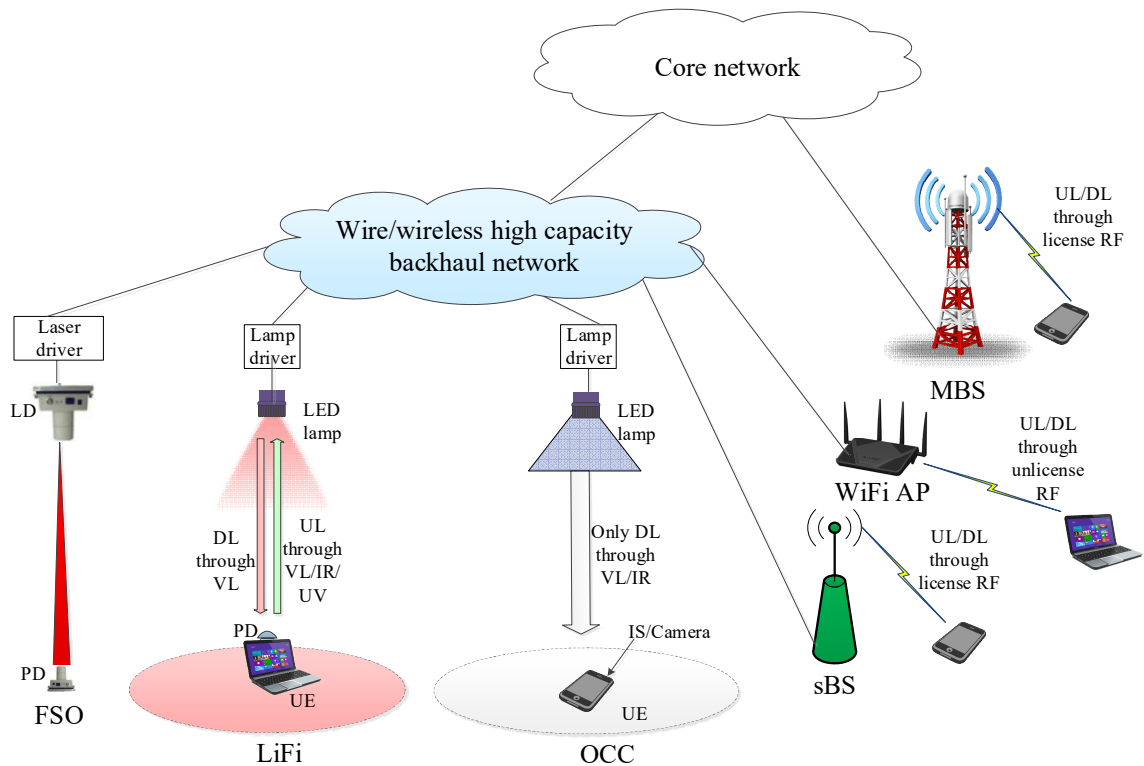

Fig. 1. Basic connectivity in different type of network technologies used for hybrid systems.

The optical/RF type of hybrid network systems can be designed in various form. Typically the LiFi or VLC system can be used for same hybrid architecture. The most important difference between LiFi and VLC system are (i) LiFi must provide point-to-multi-point communication whereas point-to-multi-point communication is not mandatory for VLC (ii) VLC must use VL for communication media whereas LiFi uses VL for forward link and VL or IR for reverse link.

*2.2. Communication based on RF*

Small-cell network: Small-cell e.g.,femtocell network technology is widely deployed in subscribers' homes to provide high QoS. The small-cell base station (sBS) is small-size cellular base-station deployed in subscribers' home. sBSs are operated in the spectrum licensed for cellular service providers.

Macrocellular network: This technology is the most extensively deployed in outdoor to provide higher coverage area and higher user mobility. However, the macrocellular base station (MBS) cannot provide higher data rate connectivity.

Microwave-link: Microwave link is the transmission of information by microwave for long distance using point-to-point communication technique.

## 3. Hybrid Optical/RF Scenarios

We can plan hybrid optical/RF systems [11], [14]-[17] through various combinations as required. The optical/RF hybrid architecture can bring great opportunities in terms of traffic offload/load balancing, improving link reliability, provisioning of hyper speed data rate, multi-tier networks, improving of spectral utilization, support for smooth handover, line-of-sight (LOS)/non-line-of-sight (NLOS) communication support, omnidirectional/directional communication support, and down-link (DL)/up-link (UL) support through different networks. Fig. 2 shows one example of hybrid optical/RF system used for communications in ship. Inside the ship, we can have small coverage networks e.g., LiFi, WiFi, OCC, and femtocells as needed. Ship can communicate through satellite. This connectivity can be possible using RF as well as optical transmission media. A high-speed FSO link can be established from the land to the ship. Also, ship-to-ship communication is performed using FSO link. Few examples of such hybrid systems are briefly discussed below:

**(i) Hybrid LiFi/WiFi and VLC/WiFi**: This form of hybrid architecture is possible both for the indoor and outdoor cases. For indoor cases, LiFi or VLC can support high data rate whereas WiFi can support comparatively wider coverage for better mobility support. VLC or LiFi also suffers from interference effect [2]. For outdoor vehicular communication, this hybrid architecture can provide better QoS for vehicle-to-everything (V2X) connectivity. For some cases LiFi (or VLC) is used for DL connectivity and WiFi for UL connectivity. For the outdoor cases e.g., vehicle-to-vehicle (V2V) and vehicle-to-infrastructure (V2I) communications, the presence of two networks increases communication reliability. Thus, this hybrid system has a great prospect for load balancing, provisioning of hyper speed data rate, improving of spectral utilization, support for smooth handover, LOS/NLOS communication support, omnidirectional/directional communication support, and DL/UL support through different network.

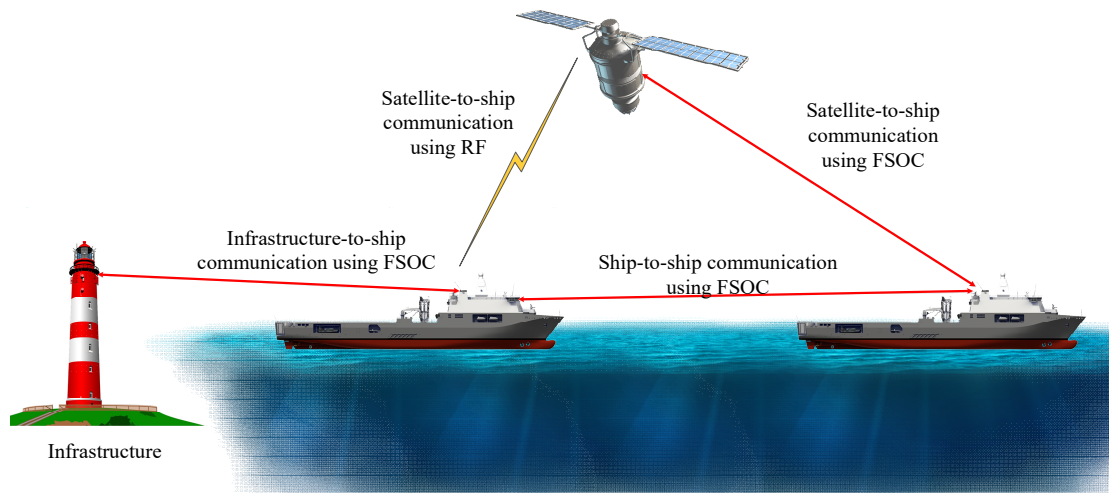

**Fig. 2.** Example of outdoor hybrid optical/RF networks for wireless connectivity with ships.

**(ii) Hybrid LiFi (or VLC)/small-cell**: The license small-cell network provides better QoS with low-to-medium mobility support but cannot support very high data rate link. Hence, this form of hybrid system can be used for indoor environment. This hybrid system offload huge traffic from small-cell networks to LiFi (or VLC) network. The coverage hole created by LiFi networks is filled by small-cell network that smoothen the handover process. Moreover, it also brings the opportunity for very high speed communication system, enlightening the spectral utilization, LOS/NLOS communication support, and omnidirectional/directional communication support.

**(iii) Hybrid FSO/RF**: The FSO system provides very high data rate point-to-point communication link in indoor as well as outdoor environment. However, FSO system heavily suffers from atmospheric effect especially in fog and dust conditions. It also suffers from transmitter-receiver alignment problem. The microwave link heavily suffers in rainy condition. Hence, this hybrid system can be used for indoor, outdoor, and even underwater communication systems for the purposes of link reliability.

**(iv) Hybrid OCC/WiFi**: Both the OCC and WiFi systems are used for localization. The localization accuracy for OCC based systems is better compared to WiFi based system. However, WiFi system provides better accuracy in NLOS conditions. Hence, this hybrid system is a good opportunity to improve the localization performance.

**(v) Hybrid LiFi/macrocell:** WiFi or small-cell is not always available in every indoor environment. The low mobility as well as comparatively lower QoS required services are connected to LiFi networks whereas higher mobility as well as comparatively higher QoS required services are connected to macrocell network in an indoor environment. Hence, this hybrid system support load balancing, hyper speed data rate, improving of spectral utilization, LOS/NLOS communication, and omnidirectional/directional communication. Moreover, the macrocell network provides the coverage for the coverage hole created by LiFi networks that smoothen the LiFi-to-LiFi handover.

## 4. Hybrid Optical/Optical Scenarios

The hybrid optical/optical systems are planed to increase the link reliability and to satisfy user QoS level. Few examples of such hybrid systems are hybrid LiFi/OCC, hybrid FSO/OCC, and visible/invisible. In hybrid LiFi/OCC, LiFi provides comparatively high data rate but not effective for comparatively longer distance communication compared to OCC system. Moreover, LiFi system is less immune against interference effect compared to OCC system. Hence, this hybrid system is an effective approach for optical wireless deployment. Hybrid FSO/OCC networks are potential for V2X communications, wherein FSO provides high speed point-to-point link and OCC is less immune against outdoor atmospheric conditions. The performance of FSO system is seriously degraded on fog and dust conditions. Therefore, this FSO/OCC hybrid system will be very effective approach for V2X communications. In visible/invisible hybrid system, communicating using optical spectrum consists of both human visible and invisible parts. Visible lights are not or limitedly provided by street light and car font/back lights during day time. Hence, for V2X communications during day time, invisible light e.g., IR can be used whereas for night time, visible light can be used. Therefore, a hybrid visible/invisible system can improve the performance in V2X communication.

## 5. Challenges and Research Direction

**Network selection**: For the hybrid system, an effective network selection technique is very essential. The network selection criteria is different for optical wireless networks compared to existing RF based networks. Hence, existing network selection technique should be modified.

**Heterogeneous receiver type**: Both the receivers for two different networks of the hybrid system should be active simultaneously. The characteristics of RF based receiver and optical based receiver are different. Hence, this is a challenging issue for hybrid network.

**Handover:** Handover is an important issue in hybrid system. Appropriate handover decision criteria and algorithm are still open research issue for optical wireless based hybrid networks.

**High-capacity backhaul network**: One of the main objectives of hybrid network systems is to provide load balancing facility. Hence, a huge amount of overall data throughput in the access network is increased. To support this data throughput, very high capacity backhaul network is essential that is still an open research issue.

## 6. Conclusions

Recently, different OWC technologies become very important part of wireless communication system. Hybrid systems comprising of optical wireless system can overcome many limitations of either RF or optical wireless based single network. This paper addresses key research issues for optical wireless hybrid networks. The hybrid architecture scenarios and their opportunities are discussed. Hence, this paper would be a good direction for future research on OWC.

## Acknowledgement

This work was supported by the MSIT (Ministry of Science and ICT), Korea, under the ITRC (Information Technology Research Center) support program (IITP-2018-0-01396) supervised by the IITP (Institute for Information & communications Technology Promotion) and the Korea Research Fellowship Program (2016H1D3A1938180).